# Novel design strategies for modulating conductive stretchable system response based on periodic assemblies

Louis Martin-Monier [1], Pierre-Luc Piveteau[1], Fabien Sorin[1]*

[1] Laboratory for Photonic and Fiber devices, Ecole Polytechnique Fédérale de Lausanne (EPFL), Lausanne, Switzerland

*Correspondence to: fabien.sorin@epfl.ch

Abstract: Soft electronics have recently gathered considerable interest thanks to their bio-mechanical compatibility. An important feature of such deformable conductors is their electrical response to strain. While development of stretchable materials with high gauge factors has attracted considerable attention, there is a growing need for stretchable conductors whose response to deformation can be accurately engineered to provide arbitrary resistance-strain relationships. The rare studies addressing this issue have focused on deterministic geometries of single rigid materials, limiting the scope of such strategies. Herein, we introduce the novel concept of periodic stretchable patterns combining multiple conductive materials to produce tailored responses. Using shortest-path algorithms, we establish a computationally efficient selection method to obtain required resistance-strain relationship. Using this algorithm, we identify and experimentally demonstrate constant resistance-strain responses up to 50% elongation using a single micro-textured material. Finally, we demonstrate counter-intuitive sinusoidal responses by integrating three materials, with interesting applications in sensing and soft robotics.

## Introduction

The integration of deformable yet conductive materials have paved the way for a new generation of truly conformable devices that can be seamlessly integrated over biological tissues. The applications of such stretchable conductive materials and structures include sensing[1,2,3], but also actuation in soft robotics[4,5,6] and stretchable optics[2,7,8]. Thus far, a large focus has been placed on maximizing system response to deformation, such as deformation sensors using deterministic geometries for demultiplied sensitivity[9,10]or development of extremely soft yet conductive hydrogels[10,11,12] in soft robotics. In specific situations however, it may be desirable to engineer a precise response to external deformation. This is the case of stretchable interconnects, which interface different rigid chips within a stretchable matrix. Given that many electrical devices such as transistors or amplifiers rely on constant current sources to function, stretchable interconnects would benefit from minimal current variation with deformation to connect these various chips under strain. In a broader manner, providing arbitrary resistance response to external deformation remains an open problem.

Microstructure offers the possibility to tailor the response in resistance with strain beyond classical deformation theory of homogeneous materials. Atypical resistive behavior with strain has been observed in various systems near-percolation composites, such as carbon-nanotube[13, 14] or silver nanowire[15,16]-based composites. Until now however, works investigating such systems have remained focused on randomly distributed networks[17], while the leverage over resistance-strain behavior has remained limited to composite loading. The lack of order at the meso- or macroscopic scale inevitably limits the variety of resistance response. Responses with negative Gauge factor have also been observed in micro or nano-structured liquid metal thin films, but the inherent randomness of the structure limits again tuning possibilities[18,19]. Recent works using deterministic geometries of folded lines of a single rigid metal, semiconductors, or rigid composite[20,21] have achieved constant resistance-strain relationships. However, such an approach provides no tunability and remains fragile, thus limited in strain. Moreover, it induces inevitable challenges to rigorously identify the elastic and plastic mechanics of these composite materials.

While recent studies have pointed to the role of orientation in 1D nanomaterials to explain complex resistance response with strain[17], no analytical approach has been proposed to understand the link between microstructure and resistance-strain behavior in ordered 2D-systems.

In this work, we propose a new methodology introducing periodic arrays in stretchable conductive systems to engineer pre-defined resistance-strain relationships. We first develop a code to accurately calculate the resistance of an NxN grid of 2D resistors with two distinct conductivities. Using this model, we show that the effective medium approach is inappropriate for dense periodic arrays with increasing conductivity contrast in stretchable electrical systems. We then demonstrate how, by fine-tuning the periodic arrangement and the choice of materials, one can design resistance-strain relationships with completely tunable Gauge factor from positive to negative passing through zero, which has particularly interesting implications for stable stretchable interconnects. To limit computing cost, we further introduce the concept of shortest resistive path to understand the evolution of resistance during elongation. Given an initial set of microstructures and materials, this approach allows in particular to provide at reduced cost the best combination of materials and microstructure to achieve a targeted response. Using an optimization algorithm, we identify by simulation the best microstructures to provide a constant resistance over extensions up to 50%. We further experimentally demonstrate such microstructures (deviation <3%) using a single micro-textured conductive stretchable material. Finally, by integrating three distinct set of materials, we discuss the possibility to deviate from purely linear responses and demonstrate counterintuitive sinusoidal resistance-strain responses, which could find interesting applications in sensing and soft robotics.

## Results and Discussion

*Breakdown of the effective medium approach in near-percolation bi-conductive systems*

Continuum mechanics provides a well-established framework linking resistance of a homogeneous material with deformation through the Gauge Factor G[22]:

$$G = \frac{\Delta R/R_0}{\Delta L/L_0} = 1 + 2\nu + \frac{\Delta\rho/\rho_0}{\varepsilon} \qquad \text{(i)}$$

Where $\nu$ is the Poisson ratio $\varepsilon$ is the strain, $\rho_0$ the initial material resistivity, $\Delta\rho$ the material relative change in resistivity. The first contribution to the Gauge Factor, also known as the geometrical contribution $(1 + 2\nu)$, is fixed by the material's properties. Although the Poisson ratio can vary largely between materials, common elastomeric materials that allow large reversible deformations are typically incompressible ($\nu \sim 0.5$ )[23], limiting the tunability of the geometrical term. The second contribution ($\frac{\Delta\rho/\rho_0}{\varepsilon}$) accounts for piezoresistive effects, and can become dominant in some materials (as high as 200 in p-type [110] single crystalline silicon[24, 25]). This term is also strongly dependent on the choice of materials, and usually occurs in systems that cannot sustain reversibly large deformations.[24,26,27]

The approach detailed above is commonly used for materials that can be considered homogeneous. Let us now consider a composite that blends two different materials (called materials A and B). This material is hence heterogeneous, typically with small inclusions of material A dispersed within a matrix of material B. To assess the resistance of such a material under deformation, one could resort to the previous homogeneous framework by considering the effective resistance of the composite $R_{eff}$, i.e. the resistance of each material pondered by their respective share of the total composite volume:

$$R_{eff}(\varepsilon) = X_A R_A(\varepsilon) + X_B R_B(\varepsilon) \qquad \text{(ii)}$$

Where $X_A$ and $X_B$ are the volume fractions of materials A and B respectively, $R_A$ and $R_B$ are the resistance associated to materials A and B. The effective medium approach is however no longer appropriate when the small inclusions of materials increase in size and the system approaches the percolation threshold. To assess the accurate resistance (called equivalent resistance $R_{eq}$ in the rest of this work) in 2D composites, we develop a model which evaluates the resistance of a random 2D resistor network made of two distinct types of resistors (total size N x N). Kirchoff's rule states that, for every site between four resistors, the sum of the currents is zero, defining a solvable linear set of $N^2$-1 equations. By injecting and extracting a known current at predefined locations, the resistance of the network between these two points can be evaluated. Given a fixed microstructure, the 2D resistor network hence allows one to calculate the accurate resistance-strain relationship for arbitrarily complex grids of reasonable size using direct matrix inversion method (see Figure 1(a)). We define material A as the background low conductivity material (in dark red) and material B as the isolated squares of high conductivity (in light red).

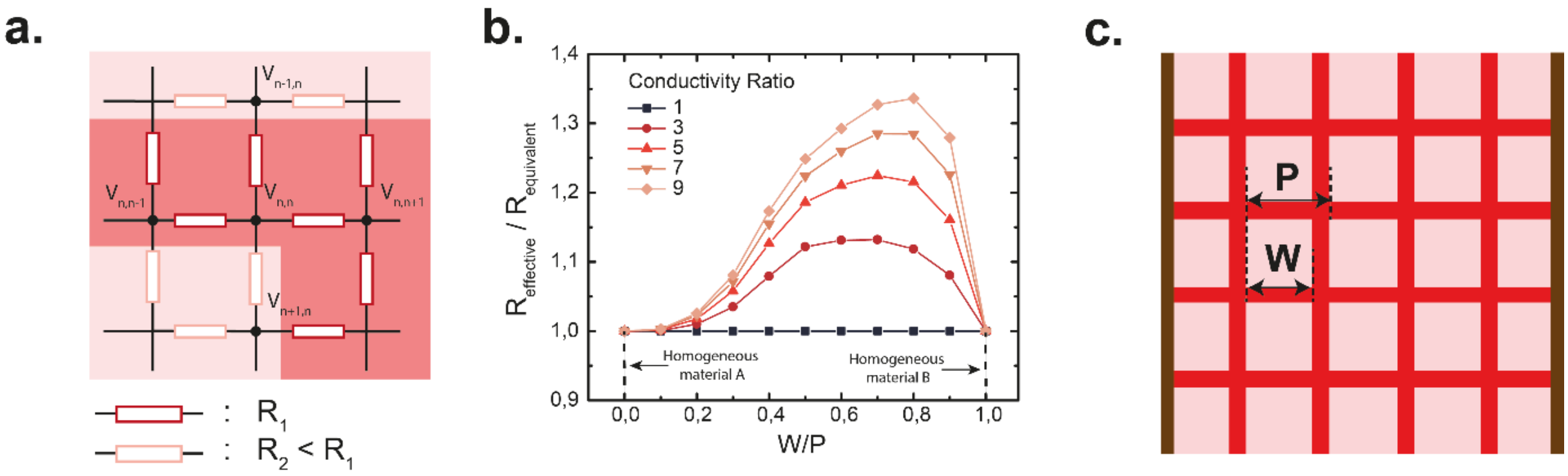

*Figure 1 – Breakdown of effective medium approach near percolation threshold in bi-conductive materials. (a) Schematic illustrating the 2D resistor network approach to model a bi-conductive material. The network is composed of resistors $R_1$ and $R_2 < R_1$ (b) Ratio of effective resistance (effective medium theory) over the equivalent resistance (exact value of network using matrix inversion method) versus the fill ratio W/P. (c) Schematic of the bi-conductive structure, indicating the values of W (particle width) and P (period). The brown lines on the edges of the sample indicate the contacts for resistance measurement.*

To evaluate the deviation of $R_{eff}$ with regards to $R_{eq}$, we now consider a heterogeneous film microstructure composed of a periodic square array of material A with conductivity $\sigma_A$ onto a substrate of material B with a conductivity $\sigma_B$ (see Figure 1(b) and 1(c)), without any elongation at this point. The square width W is gradually increased while the period P is kept fixed. Contacts of infinite conductivity are applied on the two edges of the network to calculate the equivalent resistance $R_{eq}$. For a fixed conductivity ratio $\gamma = \frac{\sigma_A}{\sigma_B} > 1$, an increase in the ratio W/P leads to a stronger deviation between effective resistance $R_{eff}$ and actual resistance $R_{eq}$ until W/P=0.8, where a maximum is reached (see Figure 1(b)). Keeping a fixed ratio W/P = 0.8, an increase in the conductivity ratio from 1 to 9 amplifies the deviation between $R_{eff}$ and $R_{eq}$ up to 35%. This deviation highlights the inadequacy of the effective medium approach in composite systems near percolation. For the rest of this work, we focus on direct evaluation of equivalent resistance to accurately determine the resistance of 2D resistor networks. We also neglect any piezo-resistive behavior for both materials given the strains involved.

### *2D resistor networks for accurate response calculation under strain*

Next, using the introduced model, we study the resistance evolution of the 2D composite under tensile deformation. We first focus on the resistance-strain relationship *parallel to the tensile axis.* For the periodic square array shown in Figure 2(a)(i), an increase in conductivity ratio essentially induces negligible change in the resistance-strain curve. Adapting the structure with particular order can change this behavior in counterintuitive ways. By introducing a shift in the structure (Figure 2(b)-(c)) along the elongation direction, the observed response deviates from the linearly increasing resistance-strain relationship with an increase in conductivity ratio. Interestingly, this behavior is not present when the periodic shift is introduced perpendicular to the elongation direction. This evolution of resistance can be understood through a simple compromise. As the sample is deformed, high conductivity regions become more separated *along* the tensile axis, creating more resistive current pathways. This effect is counterbalanced by the compression *perpendicular* to the stretching direction, which creates additional conductive current pathways (see Figure 3(a) for a detailed illustration). The interplay between these two competing phenomena define the actual global resistance change *parallel* to the tensile axis. Meanwhile, the resistance *perpendicular* to the tensile direction is exclusively driven by the reduction in distance transversally between high conductivity region, accounting for the continuous decrease in resistance. Varying the conductivity ratio can allow for Gauge factors switching at will from negative to positive passing through zero for linear arrays (Figure 2(b)-(c)). We hence identify a particular set of structures whose resistance is insensitive to strain under extended elongation. This is a particularly relevant point for stable yet stretchable electrical interconnects that can provide constant current (deviation <3%) to interfacing chips within a stretchable matrix.

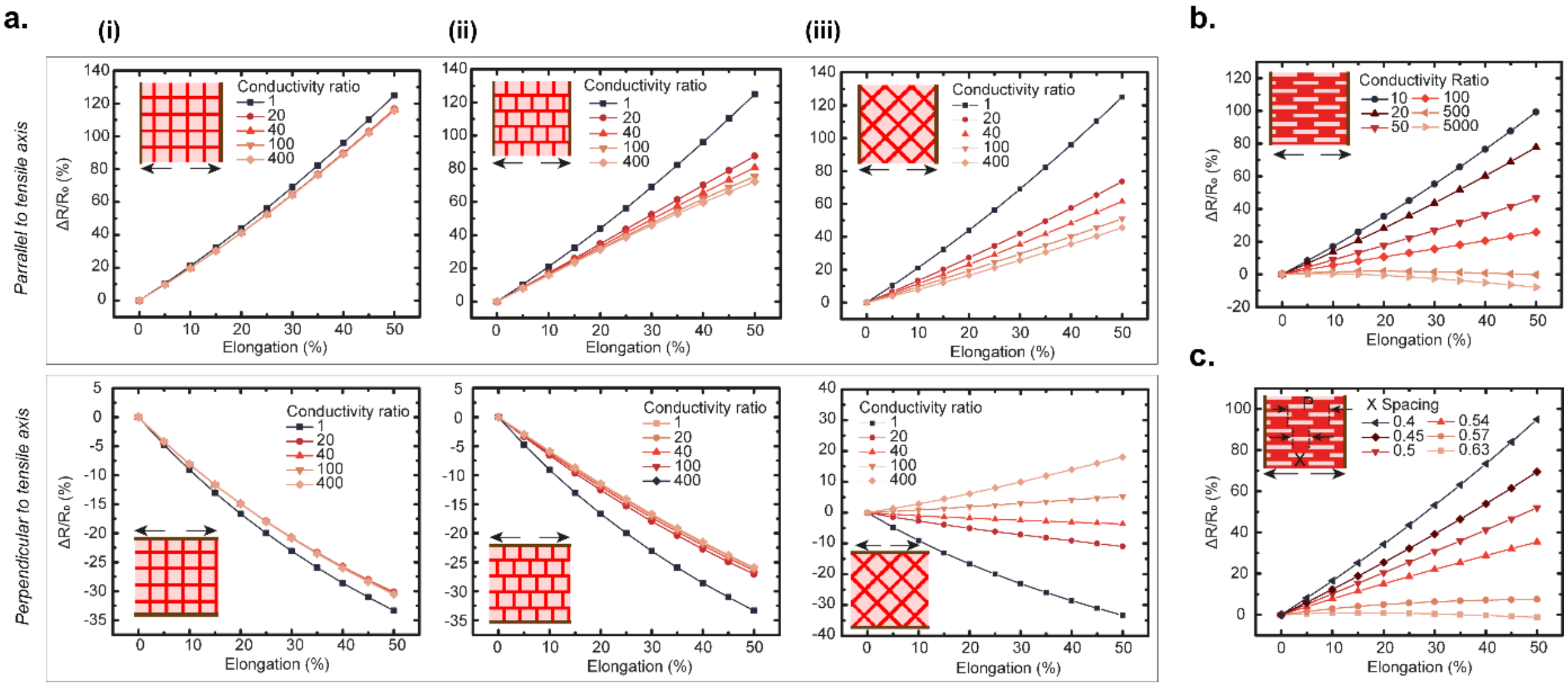

*Figure 2 – Influence of structure on electro-mechanical response. (a)-(c) Relative resistance change as a function of strain for the respective bi-conductive structures shown in inset using direct calculation method. The deformation directions is indicated by arrows, and the contacts (a) Investigation of three distinct structures with an identical highly conductive square element: (i) regular square array, (ii) shifted square array and (iii) rotated square array. The light red (resp. dark red) in the inset schematic denotes the high (resp. low) conductivity regions. Top figures correspond to stretching parallel to tensile strain, while bottom figures correspond to the associated compression perpendicular to the elongation axis. (b) Electromechanical response of periodic line array with an increase in conductivity ratio γ from 5 to 5000 (c) Electromechanical response of a periodic line array with a change in microstructure (variation in spacing X between periodic lines). Lines are of fixed length and width, and conductivity ratio is set at 500.*

*The shortest path approach*

The reverse problem, i.e. determining the proper combination of materials and microstructures to yield a desired resistance-strain response, is a relevant technological problem that requires a link between microstructural parameters, strain and resistivity. Given a library of materials and microstructures (discretized in a mesh of size NxN), a first brute force approach could consist in scanning through the various combinations of the library elements using the direct resistance calculation method detailed above. This approach may however prove computationally expensive given the $N^4$ complexity associated to matrix inversion. An alternative approach consists in simplifying the 2D resistor model into a reduced set of privileged resistor paths, corresponding to the *shortest resistive paths* for the current.

Let us now consider a microstructure where all the electrical current flows through a reduced set of shortest paths (see Figures 3(a)-(b)). Using Yen's shortest path algorithm[28], we can associate a cost to each shortest path, which helps discriminate which paths contribute most to current flow. In the particular case of the structure shown in Figures 3(a)-(b), two Paths (named Path 1 and Path 2), minimize the resistance to current at 0% strain. Additional shortest paths all show costs superior by a factor of at least 2, and are therefore further neglected. Paths 1 and 2 determined by Yen's algorithm include: (i) a first shortest path set (named Path 1, in green), made of the straight lines connecting the two contacts through the regions of lowest conductivity; (ii) a second shortest path set (named Path 2, in blue), made of the paths connecting transversally consecutives lowest conductivity regions. We now make the assumption that each path acts as a *closed channel* for current: current can only flow in or out at the extremities, i.e. the contacts. In this case, an equivalent resistance associated to each path $R_{Path\ 1}$ (resp. $R_{Path\ 2}$) can be evaluated by treating all consecutive resistor elements as a series assembly. By distinguishing elementary pathways situated parallel (//) or perpendicular (⊥) to the applied strain, a direct link between microstructural parameters and strain appears:

$$R_{Path\ j} = \sum_{Path\ j} R_i = \sum_{\substack{i\,\epsilon\, Path\ j \\ i\,\epsilon\, \perp}} f_i^{\perp}(\rho_i, L_0, S_0, \varepsilon, \nu, G_i^{\perp}) + \sum_{\substack{i\,\epsilon\, Path\ j \\ i\,\epsilon\, //}} f_i^{//}\left(\rho_i, L_0, S_0, \varepsilon, \nu, G_i^{//}\right) \quad \text{(iii)}$$

Where $\rho_i$ represents the resistivity of the i-th element of Path j, $L_0$ its length along the current direction, $S_0$ its cross section perpendicular to the current direction, ε the deformation along the tensile deformation axis, and ν the associated Poisson Ratio. $f_i^{//}$ (resp. $f_i^{\perp}$) represents the i-th elementary resistance along the shortest path parallel (resp. perpendicular) to the applied strain, and $G_i^{//} = L^{//}/W^{//}$ (resp. $G_i^{\perp} = L^{\perp}/W^{\perp}$ ) is a geometrical factor to take into account the actual width W and length L of the shortest path along the direction parallel (resp. perpendicular) to the tensile deformation (see Methods Section for full description). In the particular case of a homogeneous isotropic conductive material such as liquid metal, $f_{\perp}$ and $f_{//}$ can be explicitly expressed: $f_i^{\perp} = G_i^{\perp}.\frac{\rho_i L_0}{S_0}.(1+\varepsilon)^{1+2\nu}$ and $f_i^{//} = G_i^{//}.\frac{\rho_i L_0}{S_0}.(1+\varepsilon)^{-1}$. The range of Poisson ratio accessible using conductive elastomers defines a set of function ($f_i^{\perp}$, $f_i^{//}$ ) whose linear combination defines the full space of resistance-strain relationship attainable. The homogeneous isotropic criterium is however not validated for a number of stretchable composites, such as those based on carbon nanotubes or silver nanowires, which lead to more complex resistance-strain relationships. This indicates that the dimensionality of the resistance-strain behavior space group could be further extended.

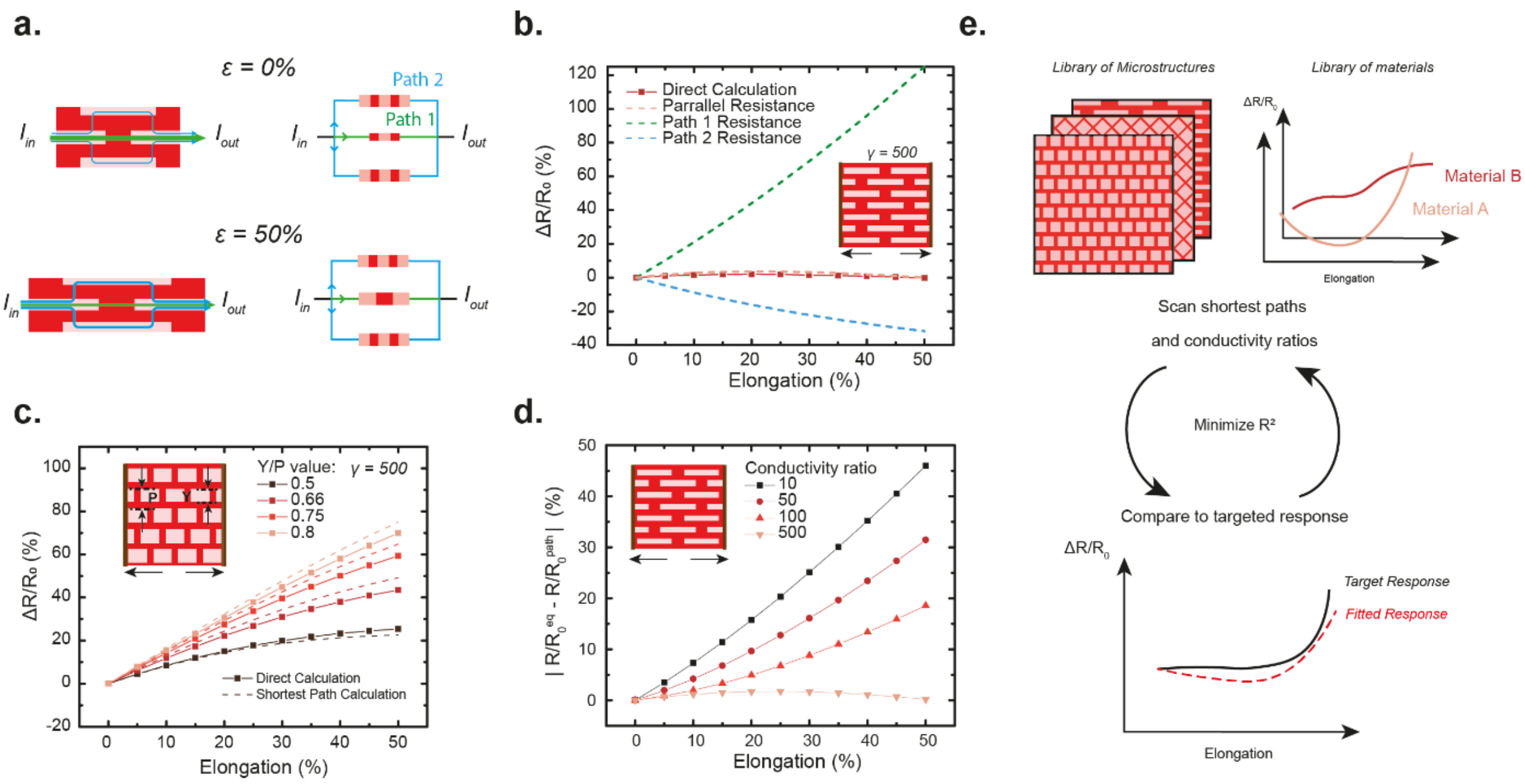

*Figure 3 – Shortest Path Method and Perspectives. (a) Schematic illustrating the different shortest paths (green and blue arrows) preferentially taken by the current under 0% and 50% elongation. (b) Calculation of resistance-strain relationship by direct calculation for the whole assembly (red dotted line) and by shortest path method at 0% strain and 50% strain (respectively green and blue dotted lines). Combining shortest path at 0% and 50% strain as a set of parallel resistor (light red dotted line) provides a good agreement with the direct calculation. All calculations are done with a fixed conductivity ratio of 500. (c) Variation of the width Y between periodic lines of fixed length and spacing. The conductivity ratio is fixed to 500. Dotted lines indicate the results obtained through shortest path method. (d) Difference between relative resistance obtained through direct calculation and shortest path method for the structure shown in inset, with a conductivity ratio increased from 10 to 500. (e) Schematic illustrating the working principle of the reverse problem: starting from an initial library of microstructures (shortest paths) and materials (resistance-strain relationship for the homogeneous material), a resulting response (labeled "fitted response" on the bottom graph) for a microstructure combining two distinct materials can be determined at low computational cost. Scanning through the possible combinations of materials and microstructure, we minimize the deviation between fitted and target response using the determination coefficient $R^2$.*

Using relation (iii), we proceed to evaluate the path resistance associated to the two shortest paths of the line array microstructure shown in Figures 3(a)-(b).

These two individual resistive paths let through a different amount of current, which varies with elongation. At 0%, both contribute at the same level to the overall sample resistance (same cost in Yen's shortest path algorithm). When stretched, the resistance associated to the second shortest path set is largely reduced, whereas the resistance associated to the first shortest path increases (see Figure 3(a), top left and bottom left). To take into account the contribution of both path sets to the total resistive behavior, each path is considered as a resistor assembled in parallel between the two contacts, with respective resistance $R_{Path\ 1}$ and $R_{Path\ 2}$. Given the hypothesis of *closed channels* made above, the resistance $R_{Path\ 1}$ and $R_{Path\ 2}$ correspond to an assembly in series of resistors along their path:

$$R_{S.P} = \left( \sum_{j=1}^{K} R_{Path\ j}^{-1} \right)^{-1} \quad \text{(iv)}$$

Figures 3(b)-(d) compare the relative equivalent resistance obtained by shortest path method and by the direct matrix inversion method. In Figure 3(b), we compare the resistance-strain relationship from direct calculation (full red line) and by the shortest path method considering (i) the first shortest path set (green dotted line), (ii) the second shortest path set (blue dotted line set) and (iii) the parallel assembly of the first and second shortest path (red dotted line) set. The good match between both methods in predicting the overall resistance-strain curve highlights the validity of the shortest path approach. In Figure 3(c), we show that the shortest path method can account properly for a changing set of structures. The predictions provided by the shortest path method are however validated for sufficiently high conductivity ratios (Figure 3(d)). This is directly related to the closed channel assumption. As highlighted in figures 3(b)-(d), this approach provides a reasonable estimation of resistance-strain relationship for a broad range of periodic structures. For structures where the shortest path at 0% and 50% strain are identical, the reasoning can more simply be limited to a combination of shortest paths at 0% strain.

*Reverse engineering of resistance-strain responses*

In figure 3(e), we detail the working principle of the reverse problem algorithm, which determines which set of microstructure and materials can yield a desired resistance strain-relationship (labeled "target response" on the bottom graph). Given an initial library of microstructures (each with a set of associated shortest paths) and materials (resistance-strain relationship for the homogeneous material), a set of resulting responses (labeled "fitted response" on the bottom graph) can be determined for each microstructure combining two distinct materials at low computational cost (linear complexity N, where N x N is the microstructure size) . Relying on the direct calculation method would quickly prove expensive given the $N^4$ complexity of the $N^2$-1 x $N^2$-1 matrix inversion method. Scanning through the possible combinations of materials and microstructure, we minimize the deviation between fitted and target response using the determination coefficient $R^2$. The accuracy of the method can further be refined using direct resistance calculation if necessary, once the shortest path approach has narrowed down the search of potential candidates. The match between fitted and target response is largely determined by the degree of freedom provided by the input materials and microstructures. In this work, we show that *linear* target responses can be accurately fitted using solely materials with a classic linear Gauge factor of 2. However, one could also envision fitting *non-linear* responses, provided that the materials library be expanded to include for instance near-percolation materials such as carbon nanotubes or silver nanowire networks.

Using this same category of materials and all the microstructures presented in Figures 1-3, we apply this selection algorithm to identify the combination providing the most constant resistance-strain response, imposing a conductivity ratio in the range 1 to 5000. The results indicate that all linear arrays (e.g. microstructures shown in Figures 2(b), 2(c), 3(b), 3(d), 4(b), 4(d)) are potentially suitable candidates, while eliminating other microstructures. Focusing now on the microstructure used in Figure 3(b)-(d), the reverse problem algorithm (shortest path pre-selection followed by direct calculation refinement) converges towards a constant resistance response with strain for an optimal conductivity ratio $\gamma$=208 using solely shortest path pre-selection, and further converges to $\gamma$=510 using the direct calculation refinement method. The following section deals with the experimental implementation of a constant resistance response based on this set of parameters (microstructure and conductivity ratio).

*Stable stretchable interconnects using a single micro-textured material*

The study has thus far focused on periodic surfaces using two different conductivities. This can straightforwardly be achieved using two materials with distinct electrical properties. This is particularly interesting given the wealth of resistance-strain relationships available for different families of stretchable conductive materials. An alternative consists in using a single stretchable conductive material but with engineered modulations in thickness, which can be equivalently seen as a system with two distinct conductivities and one identical thickness. Previous works have demonstrated how textures can be used to control sophisticated optical properties[29,30]. Inspired by these findings, we rely here on texture to tune electrical properties. Such textured systems using a single material presents two advantages: (i) it allows validation of the reverse engineering approach proposed above and (ii) it demonstrates how simple the experimental implementation can be. We proceeded by coating Eutectic Gallium-Indium (E-GaIn) alloy onto a textured Poly-dimethylsiloxane (PDMS) substrate. The textured PDMS is coated with a thin (60 nm) interfacial layer of Gold to allow for proper wetting of E-GaIn[9]. By capillarity (see Methods), the thickness of the zone between pits, labeled $t_1$, can be tailored, while the film thickness within the pits, labeled $t_2$, can be in first approximation considered constant. Gradually reducing the value of $t_1$ (or equivalently gradually increasing the conductivity ratio $\gamma$) gives rises to the wide range of resistance-deformation curves (see Figures 4(c)-(d)). Using direct resistance calculations, a fitted conductivity ratio $\gamma_{fit}$ can be associated to each thickness $t_1$. We now turn to the relative evolution of the initial resistance at 0% strain (labeled $R_0$) associated to the various conductivity ratios ($\gamma_{fit}$= 120, 475, and 1300), using (i) the experimentally measured $R_0$ and (ii) the resistance obtained through direct calculations. To facilitate comparison, results are normalized with respect to $R_0$(), as shown in Figure 4(c). The change in the initial resistance follows a similar trend between experimental and calculated data. This good overlap between experimental and simulated data highlights the validity of the proposed model. Besides from the random structures from Ref. 18, this constant resistance behavior has never been demonstrated to our knowledge at extended deformations (>~30%), and moreover with intrinsically stretchable materials.

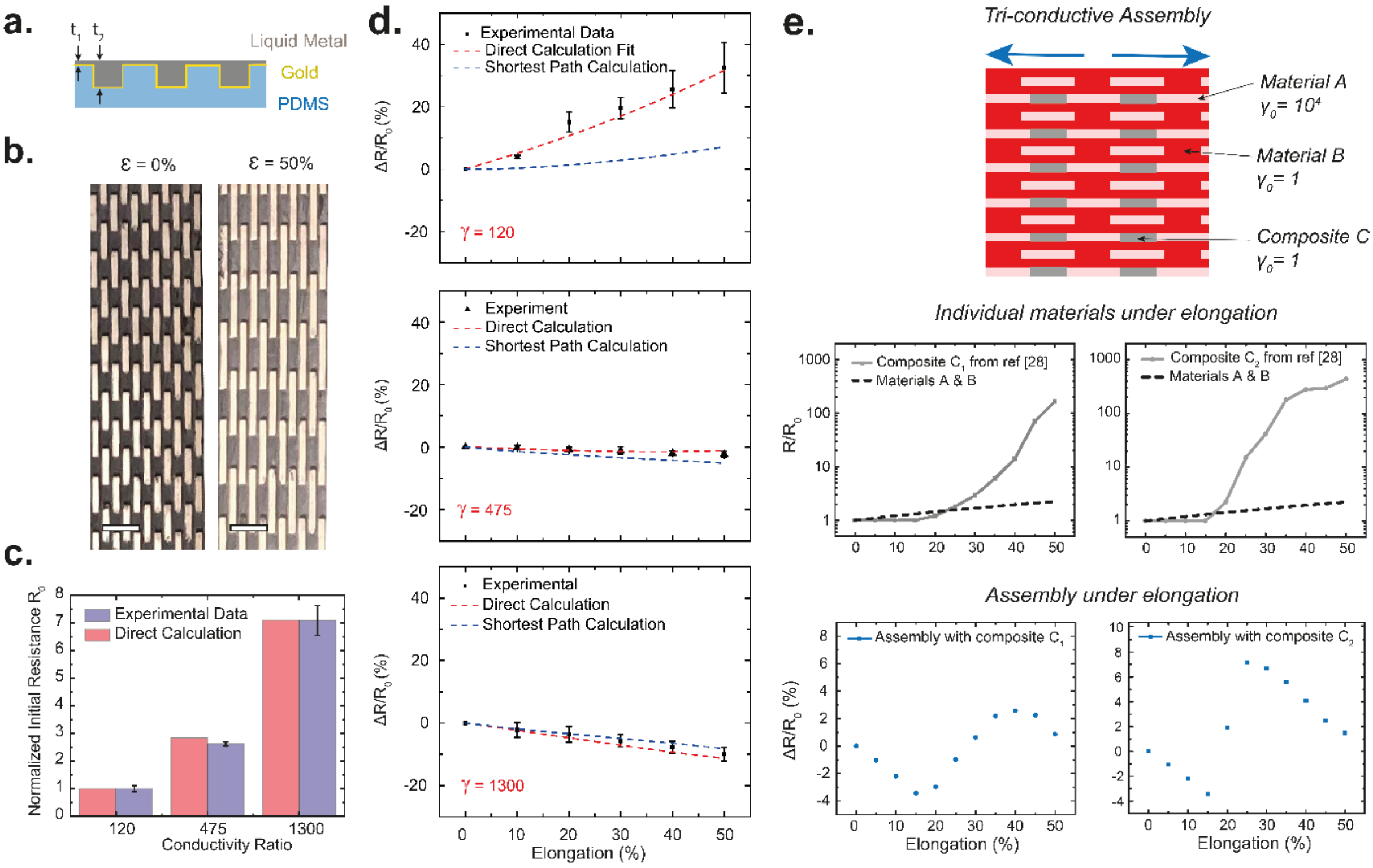

*Figure 4 – Experimental validation of the model and perspectives. (a) Schematic of the experimental sample implemented. The modulation in thickness defines an equivalent conductivity ratio between the different regions of the sample. (b) Optical Image showing the textured PDMS with liquid metal, seen from the texture side. The scale bar is 400 µm. (c) Relative change in initial resistance for different fitted conductivity ratios γ. The experimental trend observed is in line with predictions from direct calculations. (d) Relative change in resistance with strain for liquid metal films of decreasing thickness $t_1$, i.e. increasing conductivity ratio γ. Using direct calculations, a corresponding fitting conductivity ratio $\gamma_{fit}$ can be associated to each sample. Shortest Path calculations using $\gamma_{fit}$ allow for proper estimation of the experimental behavior for sufficiently large conductivity ratios. (e) (Top) Schematic of a architecture combining three materials (named A B, and Composite), with respective (relative) initial conductivities 5000, 1 and 1. (Middle) Representation of the resistance response under strain of the three materials taken individually. The materials A and B follow the traditional Ohm law with a Gauge factor of 2, while the composite material shows a strong increase in resistance typically associated to near-percolation systems. The behavior of composites 1 and 2 is based on reference [28]. (Bottom) Relative change in resistance for the corresponding assembly.*

### *Extension to three conductive stretchable materials*

To provide additional degrees of freedom, this concept of controlled electrical response in stretchable systems can be extended to more than two materials. Using near-percolation composites, we show here that resorting to three distinct materials allows for a much wider set of resulting resistance-strain curves, in particular *non-bijective* relationships. In Figure 4(e), we propose a theoretical microstructure using three materials selected from the literature materials to create a non-bijective response. We select materials A and B that both exhibit a classical linear resistance-strain behavior and a Gauge factor of 2 (Figure 4(e)(middle)). Material A is composed of Ag nanowires embedded into a poly-dimethylsiloxane (PDMS) matrix ($\gamma_{AgNW} \sim 1.8 \times 10^{3}$ S.cm$^{-1}$, based on reference [31]), while material B is made of carbon black dispersed into a PDMS matrix ($\gamma_{CB} \sim 1.7 \times 10^{-2}$ S.cm$^{-1}$, based on reference [32]). For the composite C, we select a carbon nanotube/fluoro-elastomer with a double percolated network based on reference [33] ($\gamma_{C1} \sim 1.4 \times 10^{-2}$ S.cm$^{-1}$ and $\gamma_{C2} \sim 3.3 \times 10^{-6}$ S.cm$^{-1}$, see Figure 4(e)(middle)). Using an identical thickness for materials A, B and $C_1$ provides approximately the conductivity ratios given in Figure 4(e) (respectively $10^4$, 1 and 1). $C_2$ being much less conductive, the thickness should be largely increased with respect to materials A and B so as to provide the same relative conductivity ratios. By direct matrix inversion using the 2D-resistor network model, we evaluate the evolution of such a microstructure under elongation. Interestingly, the microstructure using composite $C_1$ shows a nearly sinusoidal response (Figure 4(e)(bottom)). Similarly, composite $C_2$ shows a non-bijective response, but with a sharper transition between positive and negative values. By gradually shifting the composite's percolation threshold in consecutive linear features, one could envision a sinusoid function that would extend beyond a single period. This could open up novel applications for sensing or soft robotics, allowing for a system to come back through a given state with multiple input values simply relying on a material's microstructure.

## Conclusion

In summary, periodic electrical designs in stretchable systems create new design opportunities in stretchable electronics, including a broad range of interconnects with tunable resistance-deformation curves. With the combination of modelling based on heterogeneous 2D resistor networks and experimental measurements at different length scales, we show that the periodic microstructure and conductivity ratio determines the fundamental electrical response. Using shortest path algorithms, we show that, given an initial library of materials and microstructures, a combination of these two elements providing optimal fit with a target resistance-strain curve can be identified at low computational cost. We further show that this approach can be simply implemented experimentally using a single textured conductive material. Finally, we demonstrate by simulation the possibility to obtain non-bijective resistance-stain responses by combining up to three different materials. This study hence defines a general relationship between microstructure, mechanics, and electrical response that is broadly relevant to stretchable materials engineering.

## Acknowledgments

The authors gratefully acknowledge staff from the Center of Micro and Nanotechnology (CMi) at EPFL for help and fruitful discussion. We also thank Pr. Stephanie Lacour and Dr. Haotian Chen (EPFL-LSBI) for fruitful discussions. The authors hereby thank the financial contributors to this work, under the ERC starting grant 679211 ("Flowtonics").

## Author contribution

L.M.M had the original idea and developed the code. L.M.M and P.L.P did the experimental fabrication and characterization of the stable stretchable interconnects. L.M.M, F.S, and P.L.P wrote the manuscript.

## Competing Interests

The authors declare no competing interests.

## Methods

*Liquid metals used*

Eutectic Gallium-Indium alloy used in this work is obtained from Goodfellow, UK (Purity: 99.99%, melting point: 15.5°C).

*Master Mold for PDMS imprinting*

SU-8 resin (GM-1075, Gersteltec, Switzerland) was spin-coated at 1030 rpm for 50 seconds onto a clean Si wafer. The soft bake is done over 1 hour at 120°C using a 4°C/min ramp. The SU-8 film is then exposed using a dedicated Cr-Mask and a Mask Aligner (MA6-Gen3, Süss MicroTec, Germany), using a dose of 528 $mJ/cm^2$. Post-Exposure Bake is done for 2 hours at 90°C. Development is done using PEGMEA for 3 minutes and further rinsed using Isopropanol for 2 minutes. The resulting SU-8 pillars are silanized using Trichloro-(perfluorooctyl)silane (Sigma-Aldrich, U.S.A).

*Substrates used and treatment*

Textured Poly(dimethysiloxane) (PDMS) substrates are made by drop-casting SYLGARD 184 (Dow Chemical) over the previously described SU-8 structures, using a 10:1 monomer/curing agent ratio. The PDMS is placed under vacuum for 20 minutes before the curing step at 80°C for 24 hours.

*Stretching and fatigue tests*

All measurements were done using a four-probe measurement setup and a fixed current of 1 mA. A fixed area of 25x5 $mm^2$ (between the sensing probes) was used for all tested samples. All stretching tests were ran starting from 50% elongation and finished at 0% deformation. To gradually remove liquid metal from the substrate and hence gradually reduce thickness, a successive adsorption technique was develop. A metal syringe was passed over the metal film (side of syringe against the surface of the sample). The syringe tip could wet as well the liquid metal. The liquid metal adsorbed on the syringe after one pass was simply wiped away, removing in this way a relatively constant amount of material per pass.

*Simulation of 2D resistor network with 2 distinct conductivities*

All simulations for direct resistance calculations were done using 30 x 30 grids unless specified otherwise. All three dimensions are considered to take into account the sample's geometrical deformation. A 2D grid taking into account elementary resistivity and geometrical deformation along the 3 directions is defined. Contacts with conductivity much larger than those used in the systems are applied at the two edges of the system. Relying on this grid, we evaluate a horizontal and a vertical resistance matrix, calculating all resistance values between two neighbor elements, both along the horizontal and vertical direction. Based on Kirchoff's rule, there is an equation linking potential between a given element and its four direct neighbors. This defines a $N^2$ x $N^2$ conductance matrix $\widetilde{G_0}$, where N x N are the network dimensions. This system being overdetermined, an arbitrary point (here taken at the matrix center) is dropped from the equation system. This provides an invertible conductance sub-matrix $\tilde{G}$ of dimensions $N^2$-1 x $N^2$-1. Given a fixed input current vector $\tilde{I}$, inverting Ohm's Law provides the Potential Matrix $\tilde{U} = \tilde{G}^{-1}\tilde{I}$. Equivalently, one could fix the input voltage values and solve to determine the current since the conductance sub- matrix $\tilde{G}$ is invertible. We evaluate the resistance between two edges of the system, which can be either perpendicular or parallel to the tensile deformation. By further allowing for elementary resistor deformation, the model can predict the evolution of conductivity under tensile deformation.

*Geometrical Factor* $G_i^{\perp}/G_i^{//}$

The geometrical factor $G_i^{\perp}$ (resp. $G_i^{//}$) is a factor that accounts for the actual width and length of a path. In the simplest case where $G_i^{\perp} = G_i^{//} = 1$, the path is defined by a single line of unitary thickness linking both contacts. In this particular case, the path can simply be pictured as a linked list of elements, where each element points uniquely to the next. When $G_i^{\perp} \neq 1$ and / or $G_i^{//} \neq 1$, this picture is no longer valid. In the general case, the path can be decomposed into segments along and perpendicular to the tensile axis. For segments perpendicular (resp. parallel) to the tensile axis, geometry is taken into account by simply multiplying this particular segment resistance by the geometrical factor $G_i^{\perp} = L^{\perp}/W^{\perp}$ (resp. $G_i^{//} = L^{//}/W^{//}$), where $L^{\perp}$ (resp. $L^{//}$ ) is the corresponding segment length and $W^{\perp}$ (resp. $W^{//}$) is the corresponding segment width.

**References**

[1] A. Leber, C. Dong, R. Chandran, T. Das Gupta, N. Bartolomei, F. Sorin., Soft and stretchable liquid metal transmission lines as distributed probes of multimodal deformations. *Nat. Electron.* **3**, 316-326 (2020).

[2] Y. Qu, T. Nguyen-Dang, A. G. Page, W. Yan, T. Das Gupta, G. M. Rotaru, R. M. Rossi, V. D. Favrod, N. Bartolomei, F. Sorin, Superelastic Multimaterial Electronic and Photonic Fibers and Devices via Thermal Drawing. *Adv. Mater.* **30**, 1707251 (2018).

[3] T.Yamada, Y. Hayamizu, Y. Yamamoto, Y. Yomogida, A. Izadi-Najafabadi, D. N. Futaba, K. Hata, A stretchable carbon nanotube strain sensor for human-motion detection. *Nat. Nanotech.* **6**, 296-301 (2011).

[4] V. Cacucciolo, J. Shintake, Y. Kuwajima, S. Maeda, D. Floreano, H. Shea, Stretchable pumps for soft machines. *Nature* **572**, 516-519 (2019).

[5] J. Shintake, V. Cacucciolo, H. Shea, D. Floreano, Soft biomimetic fish robot made of dielectric elastomer actuators. *Soft Robotics 5*, 466-474 (2018).

[6] G. Agarwal, N. Besuchet, B. Audergon, J. Paik, Stretchable Materials for Robust Soft Actuators towards Assistive Wearable Devices. *Sci. Rep.* **6**, 34224 (2016).

[7] G. Mao, J. Andrews, M. Crescimanno, K. D. Singer, E. Baer, A. Hiltner, H. Song, B. Shakya, Co-extruded mechanically tunable multilayer elastomer laser. *Opt. Mat. Express*. **1**, 108-114 (2011).

[8] M. Kolle, A. Lethbridge, M. Kreysing, J. J. Baumberg, J. Aizenberg, P. Vukusic, Bio-Inspired Band-Gap Tunable Elastic Optical Multilayer Fibers. *Adv. Mater.* **25**, 2239 (2013).

[9] H. Michaud, L. Dejace, S. de Mulatier, S. P. Lacour, *2016 IEEE/RSJ International Conference on Intelligent Robots and Systems (IROS)*, 3186-91 (2016).

[10] N. Lu, C. Lu, S. Yang, J. Rogers, Highly Sensitive Skin-Mountable Strain Gauges Based Entirely on Elastomers. *Adv. Funct. Mater.* **22**, 4044 (2012).

[11] Y-J. Liu, W-T. Cao, M-G. Ma, P. Wan, Ultrasensitive Wearable Soft Strain Sensors of Conductive, Self-healing, and Elastic Hydrogels with Synergistic "Soft and Hard" Hybrid Networks. *ACS App. Mat. Inter.* **9**, 25559 (2017).

[12] K. Tian, J. Bae, S. E. Bakarich, C. Yang, R. D. Gately, G. M. Spinks, M. Panhuis, Z. Suo, J. J. Vlassak, *Adv. Mat.* **29**, 1604827 (2017).

[13] W. Luheng, D. Tianhuai, W. Peng, Influence of carbon black concentration on piezoresistivity for carbon-black-filled silicone rubber composite. *Carbon* **47**, 3151-3157 (2009).

[14] N. Hu, Y. Karube, M. Arai, T. Watanabe, C. Yan, Y. Li, Y. Liu, H. Fukunaga, Investigation on sensitivity of a polymer/carbon nanotube composite strain sensor. *Carbon* **48**, 680-687 (2010).

[15] F. Xu, Y. Zhu, Highly Conductive and Stretchable Silver Nanowire Conductors. *Adv. Mat.* **24,** 5117-5122 (2012).

[16] P. Lee, J. Lee, H. Lee, J. Yeo, S. Hong, K. H. Nam, D. Lee, S. S. Lee, S. H. Ko, Highly Stretchable and Highly Conductive Metal Electrode by Very Long Metal Nanowire Percolation Network. *Adv. Mater.* **24**, 3326-3332 (2012).

[17] L. Jin, A. Chortos, F. Lian, E. Pop, C. Linder, Z. Bao, W. Cai, Microstructural origin of resistance–strain hysteresis in carbon nanotube thin film conductors. *Proc. Natl. Ac. Sci.* **115**, 1986-1991 (2018).

[18] L. Martin-Monier, T. Das Gupta, W. Yan, S. Lacour, F. Sorin, Nanoscale Controlled Oxidation of Liquid Metals for Stretchable Electronics and Photonics. *Adv. Funct. Mater.* 2006711 (2020).

[19] C. Thrasher, Z. Farrell, N. Morris, C. Willey, C. Tabor, Mechanoresponsive Polymerized Liquid Metal Networks. *Adv. Mat.* **20**, 1903864 (2019).

[20] J. A. Fan, W.-H. Yeo, Y. Su, Y. Hattori, W. Lee, S.-Y. Jung, Y. Zhang, Z. Liu, H. Cheng, L. Falgout, M. Bajema, T. Colema, D. Gregoire, R.J. Larsen, Y. Huang, J. A. Rogers, Fractal design concepts for stretchable electronics. *Nat. Comm.* **5**, 3266 (2014).

[21] O.A. Araromi, A. Oluwaseun, A.Moritz, A. Graule, K, L. Dorsey, S. Castellanos, J. R. Foster, W.-H. Hsu, A.E. Passy, J.J. Vlassak, J.C.Weaver, C.J.Walsh, R.J.Wood, Ultra-sensitive and resilient compliant strain gauges for soft machines. *Nature* **587**, 219-224 (2020).

[22] T.G Beckwith, N. L. Buck, R.D. Marangoni, *Mechanical Measurements*. Third Edition. Addison-Wesley Publishing Co. (1982).

[23] R.H. Pritchard, P. Lava, D. Debruyne, E. M. Terentjev, Precise determination of the Poisson ratio in soft materials with 2D digital image correlation. *Soft Matter* **9**, 6037-6045 (2013).

[24] C.S. Smith, Piezoresistance Effect in Germanium and Silicon. *Phys. Rev.* **94**, 42-49 (1954).

[25] S. Middlehoek. *Silicon Sensors*. Delft, The Netherlands: Delft University Press, (1994).

[26] R. He, P. Yang, Giant piezoresistance effect in silicon nanowires. *Nat. Nanotech.* **1**, 42 (2006).

[27] A.C.H. Rowe, A. Donoso-Barrera, C. Renner, S. Arscott, Giant Room-Temperature Piezoresistance in a Metal-Silicon Hybrid Structure. *Phys. Rev. Lett.* **100**, 145501 (2008).

[28] J.Y. Yen, Finding the K Shortest Loopless Paths in a Network. *Manag. Sci.* **17**, 712-16 (1971).

[29] T. Das Gupta, L. Martin-Monier, W. Yan, A. Le Bris, T. Nguyen-Dang, A. G. Page, K-Ting Ho, F. Yesilköy, H.Altug, Y. Qu, F. Sorin, Self-Assembly of Nanostructured Glass Metasurfaces via Templated Fluid Instabilities. *Nat. Nanotech.* **14**, 320-327 (2019).

[30] A. Le Bris, F. Maloum, J. Teisseire, F. Sorin. Self-organized ordered silver nanoparticle arrays obtained by solid state dewetting . *App. Phys. Lett.* **105**, 203102 (2014).

[31] S. Zhang, H. Zhang, G. Yao, F. Liao, M. Gao, Z. Huang, K. Li, Y. Lin, Highly stretchable, sensitive, and flexible strain sensors based on silver nanoparticles/carbon nanotubes composites. *J. Alloys Compd.* **652**, 48-54 (2015).

[32] J. Shintake, E. Piskarev, S. H. Jeong, D. Floreano, Ultrastretchable Strain Sensors Using Carbon Black-Filled Elastomer Composites and Comparison of Capacitive Versus Resistive Sensors. *Adv. Mat. Tech.* **3**, 1700284 (2018).

[33] S. Shajari, M. Mahmoodi, M. Rajabian, K. Karan, U. Sundararaj, L.J. Sudak, Highly Sensitive and Stretchable Carbon Nanotube/Fluoroelastomer Nanocomposite with a Double-Percolated Network for Wearable Electronics. *Adv. Elec. Mater.* **6**, 1901067 (2020).